\documentclass[a4paper,twocolumn,superscriptaddress,pre, aps, floatfix ]{revtex4-1}
\usepackage{latexsym,amsmath}
\usepackage{amsbsy}
\usepackage[pdftex]{graphicx}
\usepackage{amssymb}
\usepackage{epstopdf}
\usepackage{verbatim}
\usepackage{physics}

\usepackage[usenames]{color}
\usepackage{capt-of} 
\usepackage{float}
\usepackage{ esint }
\usepackage{hyperref}
\usepackage{graphicx}
\usepackage[normalem]{ulem}

\usepackage{xcolor}

\begin{document}

\title{Statistical mechanics of directed networks}

\author{Mari\'an Bogu\~n\'a}
\email{marian.boguna@ub.edu}
\affiliation{Departament de F\'isica de la Mat\`eria Condensada, Universitat de Barcelona, Mart\'i i Franqu\`es 1, E-08028 Barcelona, Spain}
\affiliation{Universitat de Barcelona Institute of Complex Systems (UBICS), Barcelona, Spain}
\author{M. \'Angeles \surname{Serrano}}
\email{marian.serrano@ub.edu}
\affiliation{Departament de F\'isica de la Mat\`eria Condensada, Universitat de Barcelona, Mart\'i i Franqu\`es 1, E-08028 Barcelona, Spain}
\affiliation{Universitat de Barcelona Institute of Complex Systems (UBICS), Barcelona, Spain}
\affiliation{Instituci\'o Catalana de Recerca i Estudis Avan\c{c}ats (ICREA), Passeig Llu\'is Companys 23, E-08010 Barcelona, Spain}

\begin{abstract}
Directed networks are essential for representing complex systems, capturing the asymmetry of interactions in fields such as neuroscience, transportation, and social networks. Directionality reveals how influence, information, or resources flow within a network, fundamentally shaping the behavior of dynamical processes and distinguishing directed networks from their undirected counterparts. Robust null models are crucial for identifying meaningful patterns in these representations, yet designing models that preserve key features remains a significant challenge. One such critical feature is reciprocity, which reflects the balance of bidirectional interactions in directed networks and provides insights into the underlying structural and dynamical principles that shape their connectivity. This paper introduces a statistical mechanics framework for directed networks, modeling them as ensembles of interacting fermions. By controlling reciprocity and other network properties, our formalism offers a principled approach to analyzing directed network structures and dynamics, introducing a new perspectives, and models and analytical tools for empirical studies.
\end{abstract}

\maketitle

\section{Introduction}
A directed network~\cite{newman2018networks} is a representation of a complex system that captures the asymmetry of interactions between its elements~\cite{bianconi2008local,asllani2018structure}. Directionality enriches network structure~\cite{boguna2005generalized,angeles2007interfaces}, and is essential for understanding how influence, information, or resources flow through a system~\cite{serrano2007patterns,serrano2008structural}, fundamentally distinguishing directed networks from undirected ones. This is critical across a wide range of domains, including neuronal systems, biological processes, transportation systems, and social networks. Moreover, directionality fundamentally influences the behavior of dynamical processes on networks ~\cite{serrano2009conservation,asllani2014theory,muolo2020synchronization,PhysRevE.110.034313}. 

To gain a deeper understanding of the principles shaping real directed networks, it is crucial to define models that accurately capture their essential characteristics and organization. In general, network models enable researchers to distinguish meaningful patterns from random fluctuations and provide principled explanations for the observed regularities. The family of network models derived by maximizing the entropy of graph ensembles subject to the constraints imposed by observations in real-world networks offer the least biased prediction for their properties~\cite{park2004statistical,bianconi2009entropy}. However, designing maximum entropy models for directed networks is a challenging task. This difficulty arises from the need to account for the interplay between local node properties and global network structures.

Specifically, key features in directed networks are in-degrees and out-degrees, accounting for the number of incoming and outgoing connected neighbors, their correlations, and reciprocity~\cite{park2004statistical,garlaschelli2004patterns}, or the tendency of pairs of nodes to form bidirectional connections. Reciprocity reflects the balance or imbalance of mutual interactions and serves as a critical indicator of the underlying structural and dynamical rules governing the system. Another key properties is clustering, the tendency of pairs of neighbors to be connected, forming triangles in the network topology. In directed networks, triangles become multifaceted, splitting in seven distinct triangle motifs depending on the orientation of the arrows~\cite{fagiolo2007clustering,ahnert2008clustering}. Despite the recent introduction of a directed network model~\cite{allard2024geometric} that explains many features simultaneously in directed networks, such as reciprocity, clustering, and other structural properties, a general theoretical approach based on the maximum entropy principle is still lacking.

In this paper, we introduce a statistical mechanics framework for directed networks, treating them as systems of interacting fermions. This approach leverages concepts from quantum statistics to describe directed networks in terms of ensembles, where network connections or fermions are constrained by conserved quantities and the entropy of the ensemble is maximized to fix its probability. By framing directed networks in this way, we provide a powerful theoretical tool for modeling their structure. Our framework not only offers new insights into the organization of real-world directed networks but also provides a principled basis for constructing models that respect key empirical properties. 

\section{General formalism}

The standard approach in network science treats the nodes of a network as the fundamental units of the system, with links representing the interactions between these units. This perspective naturally aligns with real-world systems, where nodes correspond to defined entities---countries in the world trade web, proteins or genes in biomolecular interaction networks, individuals in society, and so on---making focusing on nodes intuitive and practical. However, this node-centric viewpoint poses challenges when defining models using traditional tools from statistical mechanics, as it emphasizes the entities rather than the interactions.

In this work, we adopt a different perspective by shifting the focus from the nodes of the network to the links connecting them. In our approach, links are treated as fermionic ``particles'' that can occupy distinct energy states. The phase space of possible energy states is defined by the possible links between the $N$ nodes of the network. This perspective is particularly intriguing for two reasons. First, links in a network are unlabeled, which makes them inherently indistinguishable. Second, in a simple network without multiple connections, only one link can occupy a given state, as no two identical links can exist between the same pair of nodes. These properties naturally lead to a statistical interpretation of links in a network as an ensemble of identical and independent fermions, obeying the Fermi-Dirac statistics~\cite{park2004statistical}. By reimagining directed networks in this manner, we not only provide a novel statistical framework for describing their structure but also lay the groundwork for constructing statistically rigorous principled models that capture the fundamental constraints of directed and undirected networks alike. For instance, the fermionic mapping has been instrumental in the analytical study of different aspects of networks, from the explanation of structural correlations in scale-free networks~\cite{park2003origin,boguna2004cut} to a topological phase transition with divergent entropy involving the reorganization of network cycles~\cite{vanderkolk2022anomalous}.

\begin{figure}
	\centering
	\includegraphics[width=\linewidth]{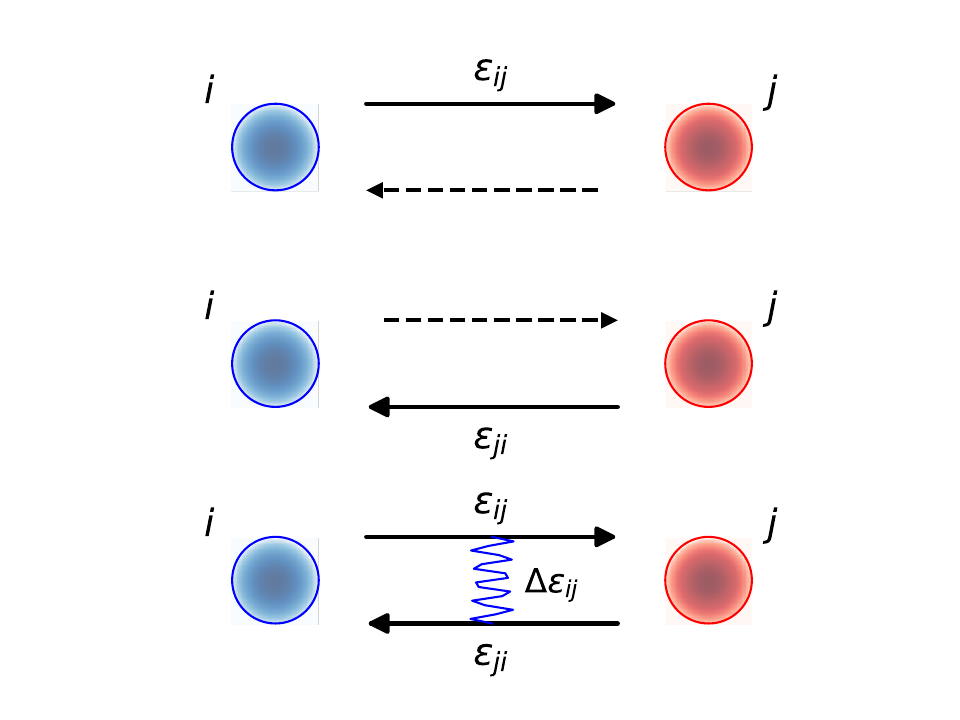}
	\caption{Possible fermionic states between a pair of nodes $i$ and $j$, and their associated energies. The solid arrow indicates the presence of a directed link and the dashed arrow an empty state. When the two fermions occupy simultaneously the two states, $ i\rightarrow j$ and $j\rightarrow i$, the total energy includes a correction $\Delta\varepsilon_{ij}$ added to the sum of the energies of the partially occupied states.} 
	\label{fig:fig1}
\end{figure}

\subsection{Fermionic approach to directed networks}
Given a pair of nodes $i$ and $j$, we define two distinct states, $ i\rightarrow j$ and $j\rightarrow i$, which can be occupied by  links, or fermions, pointing from $i$ to $j$ and from $j$ to $i$, respectively, see Fig.~\ref{fig:fig1}. Each state $i\rightarrow j$ has an associated energy $\varepsilon_{ij}$. The occupancy of these states is described by the asymmetric adjacency matrix $\{a_{ij}\}$, which equals $1$ if the state $i \rightarrow j$ is occupied and $0$ otherwise, analogous to the occupation number of states in systems of indistinguishable particles. All the topological properties of the network can be computed from the adjacency matrix. For instance, the number of incoming connections to a node, or incoming degree, is $$k_{in} = \sum_{j=1}^{N} a_{ji},$$ where $N$ is the total number of nodes in the network. Analogously, the number of outgoing connections from a node, or outgoing degree, is $$k_{out} = \sum_{j=1}^{N} a_{ij}.$$

Reciprocity implies pairs of nodes with links pointing in both directions, as shown in the sketch at the bottom of Fig.~\ref{fig:fig1}. In random network models, a certain default level of reciprocity is attained when links are independent, or fermions are non-interacting. However, higher or lower values require that links are correlated, or fermions are interacting. To account for this possibility, we assume that the energy of two links occupying the two states $i\rightarrow j$ and $j\rightarrow i$ simultaneously, that is, of mutual interactions, is $\tilde{\varepsilon}_{ij}$. In general, $\tilde{\varepsilon}_{ij}$ is different from $\varepsilon_{ij}+\varepsilon_{ji}$. 

Due to the indistinguishability of links in a network, any directed network can be represented in the Fock space using the basis $\{\ket{a}\equiv\bigotimes_{i,j}\ket{a_{ij}}\}$ defining the number of particles/links occupying the set of possible single-particle states. Thus, the representation of the Hamiltonian of the network $\hat{H}$ in the basis of the Fock space defined by the adjacency matrix is
\begin{equation}
\bra{a} \hat{H} \ket{a} = \sum_{i<j} \left[ a_{ij}\varepsilon_{ij} + a_{ji}\varepsilon_{ji} + a_{ij}a_{ji}\Delta\varepsilon_{ij} \right],
\end{equation}
where $$\Delta\varepsilon_{ij} = \tilde{\varepsilon}_{ij} - \varepsilon_{ij} - \varepsilon_{ji}$$ is the correction due to the interaction of two fermions occupying the two states $i\rightarrow j$ and $j\rightarrow i$. When $\Delta\varepsilon_{ij} > 0$, the presence of two links connecting the same pair of nodes in opposite directions is energetically unfavorable, and thus reciprocity is lower than in the random case. Conversely, when $\Delta\varepsilon_{ij} < 0$, link reciprocity is higher than random.

In analogy to the case of indistinguishable quantum particles, it is more convenient to work in the grand canonical ensemble, where the constraints are:
\begin{itemize}
\item the number of fermions (links) is fixed on average,
\item and the average energy is fixed as well.
\end{itemize}
In our formalism, this implies that the total number of links is a random variable that is fixed on average by the chemical potential $\mu$. The grand partition function of the system is given by
\begin{align}
&Z &=& \quad Tr\left( e^{-\beta (\hat{H}-\mu \hat{N}_L)} \right)\\ \nonumber
&&=& \prod_{i<j}\left( 1 + e^{-\beta(\varepsilon_{ij}-\mu)} + e^{-\beta(\varepsilon_{ji}-\mu)} + e^{-\beta(\tilde{\varepsilon}_{ij}-2\mu)} \right),
\end{align}
where $\hat{N}_L$ is the number of links operator, and the inverse temperature $\beta$ controls the average energy of the network. The chemical potential $\mu$ fixes the average in-degree (and out-degree) through the relation
\begin{equation}
\langle k_{\text{in}} \rangle = \langle k_{\text{out}} \rangle = \frac{1}{N\beta} \left(\frac{\partial \ln Z}{\partial \mu}\right)_{\beta},
\end{equation}
 and the entropy of the ensemble can be computed from the partition function as
\begin{equation}
S = \ln Z - \beta \left(\frac{\partial \ln Z}{\partial \beta}\right)_{\mu}.
\label{eq:entropy}
\end{equation}

Beyond these global thermodynamic properties, the probability of the ensemble generating a graph with adjacency matrix $\{a_{ij}\}$ is computed as the probability of a particular configuration of the system
\begin{equation}
\text{Prob}(\{ a_{ij} \}) = \frac{1}{Z} \prod_{i<j} e^{-\beta\left[(\varepsilon_{ij}-\mu)a_{ij}+(\varepsilon_{ji}-\mu)a_{ji}+a_{ij}a_{ji}\Delta\varepsilon_{ij} \right]}.
\end{equation}
The joint probability of the pair of states $i \rightarrow j$ and $i \leftarrow j$ between nodes $i$ and $j$ is
\begin{equation}
\text{Prob}(a_{ij},a_{ji}) = \frac{e^{-\beta\left[(\varepsilon_{ij}-\mu)a_{ij}+(\varepsilon_{ji}-\mu)a_{ji}+a_{ij}a_{ji}\Delta\varepsilon_{ij} \right]}}{1 + e^{-\beta(\varepsilon_{ij}-\mu)} + e^{-\beta(\varepsilon_{ji}-\mu)} + e^{-\beta(\tilde{\varepsilon}_{ij}-2\mu)}}.
\label{eq:pjoin}
\end{equation}
Finally, the probability of a directed link existing between nodes $i$ and $j$, $p_{ij} \equiv \text{Prob}(a_{ij}=1)$, is
\begin{equation}
p_{ij} = \frac{e^{-\beta(\varepsilon_{ij}-\mu)} + e^{-\beta(\tilde{\varepsilon}_{ij}-2\mu)}}{1 + e^{-\beta(\varepsilon_{ij}-\mu)} + e^{-\beta(\varepsilon_{ji}-\mu)} + e^{-\beta(\tilde{\varepsilon}_{ij}-2\mu)}}.
\label{eq:pijcorr}
\end{equation}
Equation~\eqref{eq:pijcorr} can be used to evaluate the average in- and out-degrees of individual nodes as $$\kappa_{out,i}=\sum_j p_{ij}$$ and $$\kappa_{in,i}=\sum_j p_{ji},$$ and the chemical potential as the solution of the equation 
\begin{equation}
\langle k_{\text{in}} \rangle N=\sum_{i,j,j\ne,i} p_{ij}.
\label{eq:kinmean}
\end{equation}

Finally, we can use these results to evaluate the reciprocity of the network $r$, defined as the ratio between the number of reciprocated links and the total number of links. Thus,
\begin{equation}
r = \frac{2\sum_{i<j}  p_{ij}(1,1)}{\sum_{i,j\ne i} p_{ij}},
\label{eq:r}
\end{equation}
where we have defined $p_{ij}(1,1)\equiv \text{Prob}(a_{ij}=1, a_{ji}=1)$. It is important to mention here that the freedom to chose the interaction energies $\Delta \epsilon_{ij}$ enables the possibility to adjust the level or reciprocity for particular sets of nodes, or with specific topological properties.

\subsection{Non-interacting fermions}
When the links are independent or, equivalently, the fermions are non-interacting, $\Delta \varepsilon_{ij}=0$ and the energy is $\tilde{\varepsilon}_{ij}=\varepsilon_{ij}+\varepsilon_{ji}$. In this situation, the connection probability $p_{ij}$ of a directed link between nodes $i$ and $j$ takes the simple form
\begin{equation}
p_{ij}^{ni}=\frac{1}{1+e^{\beta(\varepsilon_{ij}-\mu)}}.
\label{eq:pij}
\end{equation}
The joint probability $\text{Prob}(a_{ij},a_{ji})$ factorizes as $\text{Prob}(a_{ij},a_{ji})=p_{ij}^{ni}p_{ji}^{ni}$, and so does the partition function
\begin{equation}
Z = \prod_{i<j}\left( 1 + e^{-\beta(\varepsilon_{ij}-\mu)}\right)\left( 1 + e^{-\beta(\varepsilon_{ji}-\mu)}\right).
\end{equation}
Finally, the reciprocity becomes
\begin{equation}
r = \frac{2\sum_{i<j} p_{ij}^{ni}p_{ji}^{ni}}{\sum_{i,j \ne i} p_{ij}^{ni}},
\label{eq:reciprocity}
\end{equation}
which corresponds to the reciprocity expected by pure chance.

\subsection{Interacting fermions}
The connection probability of the system without interactions, $p_{ij}^{ni}$ in Eq.~\eqref{eq:pij}, can be used to rewrite the connection probability for a directed link in the case of interacting fermions, $p_{ij}$ in Eq.~\eqref{eq:pijcorr}, which leads to
\begin{equation}
p_{ij}=p_{ij}^{ni} \frac{1-p_{ji}^{ni}(1-e^{-\beta \Delta \varepsilon_{ij}})}{1-p_{ij}^{ni}p_{ji}^{ni}(1-e^{-\beta \Delta \varepsilon_{ij}})}.
\label{eq:pijvspij}
\end{equation}
In the case of weak interactions or high temperature, the term $\beta \Delta \varepsilon$ is small, leading to $p_{ij} \approx p_{ij}^{ni}$. Similarly, as seen from Eq.~\eqref{eq:pijvspij}, the connection probability remains unchanged by fermionic interactions in the limits $p_{ij}^{ni}\rightarrow 0$ or $p_{ij}^{ni}\rightarrow 1$, where $p_{ij}=p_{ij}^{ni}$ again holds. In these extreme situations, the lack or excess of bidirectional links leaves no room for the network to exhibit sensitivity to changes in the tendency for reciprocity. We will use this general property in the next section when dealing with specific models.

\section{Specific random network models}
So far, we have not specified the energies of the states $\{\varepsilon_{ij}\}$, which ultimately define the particular model at hand. To illustrate the power of our approach, we focus on two different models within our formalism: the non-interacting Directed Soft Configuration Model (NI-DCM)~\cite{park2004statistical}, and the non-interacting Directed Geometric Soft Configuration Model (NI-DGCM)~\cite{allard2024geometric}. Furthermore, we also derive their maximum entropy interacting counterparts (I-DCM and I-DGCM).

\subsection{Directed Configuration Model}
To derive the probability of connection of the NI-DCM~\cite{park2004statistical,kim2012constructing} within our formalism, we make the simplest assumption that the energy of a directed link connecting nodes $i$ and $j$ comes from two sources: the energetic cost that node $i$ incurs when creating an outgoing connection, $\varepsilon_{out,i}$, plus the energetic cost that node $j$ incurs when accepting an incoming connection, $\varepsilon_{in,j}$. The total energy of the fermionic state is then
\begin{equation}
\varepsilon_{ij} = \varepsilon_{out,i} + \varepsilon_{in,j}.
\label{eq:SCM}
\end{equation}
Thus, each node in the network is characterized by an associated vector $(\varepsilon_{in}, \varepsilon_{out})$ accounting for incoming and outgoing connections. The distribution of such variables is given by the probability density function $\rho(\varepsilon_{in}, \varepsilon_{out})$, with marginal distributions for $\varepsilon_{in}$ and $\varepsilon_{out}$, $\rho_{in}(\varepsilon_{in})$ and $\rho_{out}(\varepsilon_{out})$.

A priori, the formalism works for an arbitrary number of fermions between $0$ and $N(N-1)$. However, real complex networks are sparse, meaning that the average in- and out-degrees, $\langle k_{in} \rangle=\langle k_{out} \rangle$, are size-independent. In the rest of the paper, we consider ensembles of sparse networks.

\subsubsection{Non-Interacting Directed Configuration Model (NI-DCM)}
Using Eq.~\eqref{eq:kinmean}, and assuming that $\Delta \varepsilon_{ij}=0$, and replacing sums by integrals, we can write
\begin{equation}
\langle k_{in} \rangle=Nz \int \int \frac{\rho_{in}(\varepsilon_{in})\rho_{out}(\varepsilon_{out})}{z+e^{\beta \varepsilon_{in}} e^{\beta \varepsilon_{out}}}d\varepsilon_{in}d\varepsilon_{out},
\end{equation}
where we have defined the fugacity in the standard way as $z \equiv e^{\beta \mu}$. Imposing sparsity in the thermodynamic limit of this particular model implies that the fugacity must scale with the system size as $z \sim N^{-1}$. This implies that the chemical potential takes the size-dependent form
\begin{equation}
\mu = \frac{1}{\beta} \ln{
\left[
\frac{\langle k_{in} \rangle}
{N \langle e^{-\beta \varepsilon_{in}} \rangle \langle e^{-\beta \varepsilon_{out}} \rangle} 
\right]
},
\label{eq:muDCM}
\end{equation}
provided that $\langle e^{-\beta \varepsilon_{in}} \rangle$ and $\langle e^{-\beta \varepsilon_{out}} \rangle$ are bounded. In this case, the dependence between expected in- and out-degrees of nodes, $\kappa_{in}$ and $\kappa_{out}$, and the in and out energies, $\varepsilon_{in}$ and $\varepsilon_{out}$, become
\begin{equation}
\kappa_{in} = \frac{\langle k_{in} \rangle}{\langle e^{-\beta \varepsilon_{in}}\rangle} e^{-\beta \varepsilon_{in}} \; \; \mbox{and} \; \; \kappa_{out} =  \frac{\langle k_{out} \rangle}{\langle e^{-\beta \varepsilon_{out}}\rangle} e^{-\beta \varepsilon_{out}}.
\label{eq:kappas}
\end{equation}
Substituting Eq.~\eqref{eq:SCM} into Eq.~\eqref{eq:pij} and using Eq.~\eqref{eq:muDCM} and Eq.~\eqref{eq:kappas}, the connection probability in Eq.~\eqref{eq:pij} becomes the one for the directed soft configuration model:
\begin{equation}
p_{ij}^{ni} = \frac{1}{1 + \frac{\langle k_{in} \rangle N}{\kappa_{out,i} \kappa_{in,j}}}.
\label{eq:pijDCM}
\end{equation}
Notice that, when the energies of states in Eq.~\eqref{eq:SCM} are temperature-independent, the limit $\beta \rightarrow 0$ converges to the directed version of the classical Erd\"os-R\'enyi ensemble~\cite{erdo1959s} because, in this limit, the expected degree of all nodes converges to the same value, as can be seen from Eq.~\eqref{eq:kappas}. In the opposite limit, when $\beta \gg 1$, the degree distribution becomes more heavy tailed and, depending on the distribution of energies, it may undergo a phase transition to a condensed phase where a finite fraction of nodes accumulate an extensive number of links, as shown in~\cite{Bianconi:2001yq}. This effect will occur when the averages $\langle e^{-\beta \varepsilon_{in}} \rangle$ and/or $\langle e^{-\beta \varepsilon_{out}} \rangle$ diverge for $\beta> \beta_c$ for some critical inverse temperture $\beta_c$.

An alternative approach to Eq.~(\ref{eq:SCM}) is to fix the expected in- and out-degree distributions by defining temperature-dependent energy levels as
\begin{equation}
\varepsilon_{ij}=-\frac{1}{\beta} \ln{\left(\kappa_{out,i} \kappa_{in,j} \right)},
\end{equation}
and the chemical potential as
\begin{equation}
\mu=-\frac{1}{\beta} \ln{\left[\langle k_{in} \rangle N \right]}.
\end{equation}
These choices lead to the same connection probability Eq.~\eqref{eq:pijDCM}, with the difference that now the expected in- and out-degrees are temperature independent and, thus, the degree distribution is fixed. Temperature-dependent energy levels appear in strongly interacting systems~\cite{rushbrooke1940statistical,landsberg1954statistical,elcock1957temperature,de2020statistical}.

The entropy of the ensemble can be calculated using Eq.~\eqref{eq:entropy}, whose leading terms are
\begin{equation}
S=\langle k_{in} \rangle N (\ln{\left[\langle k_{in} \rangle N \right]}-1)+O(\ln{N}),
\end{equation}
recovering results in~\cite{bianconi2009entropy}. Notice that this expression does not depend on the ensemble temperature, only on the total number of links, which is a property that is fixed in the ensemble and does not depend on the degree distribution. This means that the same expression holds in the alternative definition of the model where the energy of the states is temperature-dependent. 

Finally, the reciprocity of the ensemble can be evaluated using Eq.~\eqref{eq:reciprocity}, and reads
\begin{equation}
r=\frac{\langle k_{in}k_{out}\rangle^2}{N \langle k_{in} \rangle^3}-\frac{\langle k_{in}^2k_{out}^2\rangle}{N^2 \langle k_{in} \rangle^3}
\approx \frac{\langle k_{in}k_{out}\rangle^2}{N \langle k_{in} \rangle^3}.
\label{eq.rni-scm}
\end{equation}
Thus, the reciprocity of the NI-SCM vanishes in the thermodynamic limit, even though it can become significant if the in- and out-degrees of nodes are positively correlated and their distributions heavy tailed.

\subsubsection{Interacting Directed Configuration Model (I-DCM)}
The probability for a directed link in this model can be found by substituting Eq.~\eqref{eq:SCM} into Eq.~\eqref{eq:pijcorr}, with $\Delta \varepsilon_{ij} \ne 0$, and imposing sparsity, which would lead again to Eq.~\eqref{eq:muDCM} and Eq.~\eqref{eq:kappas} when $\langle e^{-\beta \varepsilon_{in}} \rangle$ and $\langle e^{-\beta \varepsilon_{out}} \rangle$ are bounded. Alternatively, Eq.~\eqref{eq:pijvspij}, which relates the connection probabilities in the interacting and non-interacting formulations, provides a shortcut. The connection probability of the NI-DCM is size-dependent with $p_{ij}^{ni}$ scaling as $N^{-1}$, hence approaching zero in the thermodynamic limit. In this extreme, Eq.~\eqref{eq:pijvspij} indicates that $p_{ij} \approx p_{ij}^{ni}$, which implies that the energies $\varepsilon_{in}$ and $\varepsilon_{out}$, along with $\beta$ and $\mu$, define the in- and out-degree distributions as in the non-interacting model. 

In contrast, the joint probability $\text{Prob}(a_{ij},a_{ji})$ in the I-DCM does not factorize, thereby enabling to tune the reciprocity. The reciprocity can be calculated from Eq.~(\ref{eq:r}), using the probability to have a bidirectional connection between nodes $i$ and $j$ from Eq.(\ref{eq:pjoin}) after imposing that the two links are present simultaneously, $a_{ij}=a_{ji}=1$. Using that $$e^{\beta(\varepsilon_{ij}-\mu)}=\frac{N \langle k_{in} \rangle}{\kappa_{out,i} \kappa_{in,j}},$$ the reciprocity is
\begin{align}
&r=\frac{2}{N\langle k_{in} \rangle} \times \\ \nonumber
&\sum_{i<j}\frac{\frac{\kappa_{out,i} \kappa_{in,j}}{N\langle k_{in} \rangle}\frac{\kappa_{out,j} \kappa_{in,i}}{N\langle k_{in} \rangle}e^{-\beta\Delta\varepsilon_{ij}}}{1+\frac{\kappa_{out,i} \kappa_{in,j}}{N\langle k_{in} \rangle}+\frac{\kappa_{out,j} \kappa_{in,i}}{N\langle k_{in} \rangle}+\frac{\kappa_{out,i} \kappa_{in,j}\kappa_{out,j} \kappa_{in,i}}{N\langle k_{in} \rangle)^2}e^{-\beta\Delta\varepsilon_{ij}}},
\end{align}
which, to leading order in $N$, gives
\begin{equation}
r=\frac{1}{(N\langle k_{in} \rangle)^3}\sum_{i,j}\kappa_{out,i} \kappa_{in,i} \kappa_{out,j} \kappa_{in,j} e^{-\beta\Delta\varepsilon_{ij}}.
\end{equation}
This result implies that reciprocity vanishes in the thermodynamic limit. The specific form in which $r \rightarrow 0$ as $N \rightarrow \infty$ depends on the form of the interaction energy. In all cases, when $\Delta\varepsilon_{ij}>0$, reciprocity is energetically unfavorable, and thus lower than in the NI-SCM for the same temperature; conversely, when $\Delta\varepsilon_{ij}<0$, link reciprocity is higher.

For instance, a constant value independent of the specific pair of nodes, $\Delta\varepsilon_{ij}=\varepsilon$, leads to 
\begin{equation}
r=\frac{e^{-\beta\varepsilon}}{N\langle k_{in} \rangle^3}\langle k_{in}k_{out}\rangle^2,
\end{equation}
meaning that the interaction introduces a temperature-dependent rescaling as compared to the reciprocity of the NI-SCM in Eq.~(\ref{eq.rni-scm}).  

If, instead of a constant value, the nodes in the interaction have an additive contribution to the interaction correction energy, $\Delta\varepsilon_{ij}=\varepsilon_i +\varepsilon_j$, then
\begin{equation}
r=\frac{1}{N\langle k_{in} \rangle^3}\left(\sum_{i}\kappa_{out,i} \kappa_{in,i} e^{-\beta\varepsilon_{i}}\right)^2.
\end{equation}
If $\varepsilon_i$ is proportional to the temperature, $\varepsilon_i \propto 1/\beta$, the NI-SCM behavior is recovered with a temperature-independent constant rescaling. Additionally, it can incorporate dependencies on the hidden degrees of the corresponding node, for instance, $\varepsilon_i=-1/\beta \ln(\kappa_{out,i} \kappa_{in,i})$, and then
\begin{equation}
r=\frac{1}{N\langle k_{in} \rangle^3}\langle (k_{in}k_{out})^2\rangle^2.
\end{equation}
Again, local correlations between the incoming and outgoing degrees of a node control the velocity of the reciprocity's decay. The results above also imply that a size-dependent negative interaction energy with intensity $|\varepsilon| \propto 1/\beta \ln N$ could counteract the decay of reciprocity in the SCM model and produce a finite value even in the thermodynamic limit.

\subsection{Directed $\mathbb{S}^d$ Model}

As we have seen in the previous section, reciprocity vanishes in the thermodynamic limit of the DCM even when fermions interact. Similarly, clustering also vanishes due to the size dependence of the connection probability. Finite reciprocity and clustering can be obtained in geometric networks, where nodes are distributed in an underlying metric space such that a distance $x_{ij}$ can be defined between any pair of nodes~\cite{serrano2008selfsimilarity,boguna2021network,serrano2022shortest}. In this situation, we assume that the energies of sending out or accepting a link are supplemented with a cost associated with the distance between the nodes. Thus, the total energy of a link is
\begin{equation}
\varepsilon_{ij} = \varepsilon_{out,i} + \varepsilon_{in,j} + f(x_{ij}),
\label{eq:energySd}
\end{equation}
where $f(x)$ is a monotonically increasing function of the distance. An interesting choice is a logarithmic function $f(x_{ij}) = \ln{x_{ij}}$, with nodes distributed in a $d$-dimensional Euclidean space $\mathbb{R}^d$ according to a Poisson point process with constant density $\delta$. 

\subsubsection{Non-Interacting Directed $\mathbb{S}^d$ Model (NI-DSM)}
When  $\Delta \varepsilon_{ij}=0$, the expected out-degree of a node with energy $\varepsilon_{out}$ and located, without loss of generality, at the origin of coordinates is given by
\begin{equation}
\langle k_{out} (\varepsilon_{out}) \rangle = \delta \int \rho(\varepsilon_{in}) d\varepsilon_{in} \int_0^\infty \frac{V_{d-1} r^{d-1}}{1 + r^{\beta} e^{\beta(\varepsilon_{in}+\varepsilon_{out}-\mu)}} dr,
\end{equation}
where $V_{d-1} = 2\pi^{d/2}/\Gamma(d/2)$ is the volume of a $(d-1)$-sphere. This expression can be rewritten for $\beta>d$ as~\footnote{The case $\beta<d$ can be analyzed as in~\cite{boguna2020small,vanderkolk2022anomalous,Kolk:2024aa}.}
\begin{equation}
\langle k_{out} (\varepsilon_{out}) \rangle = \delta V_{d-1} I(\beta,d) \langle e^{-d\varepsilon_{in}} \rangle e^{d\mu} e^{-d\varepsilon_{out}},
\end{equation}
where
\begin{equation}
I(\beta,d) = \int_0^\infty \frac{t^{d-1} dt}{1+t^\beta} = \frac{\pi}{\beta \sin{\frac{d \pi}{\beta}}}.
\end{equation}
Thus, if we redefine the expected out- and in-degrees as $\kappa_{out} \equiv e^{-d\varepsilon_{out}}$ and $\kappa_{in} \equiv e^{-d\varepsilon_{in}}$, with $\mu = -\frac{1}{d} \ln{\left( \delta V_{d-1} I(\beta,d) \langle k_{in} \rangle \right)}$, the connection probability becomes
\begin{equation}
p_{ij} = \frac{1}{1 + \chi_{ij}^\beta} \;\; \mbox{with} \;\; \chi_{ij}\equiv \frac{x_{ij}}{(\hat{\mu} \kappa_{out,i} \kappa_{in,j})^{\frac{1}{d}}},
\end{equation}
and
\begin{equation}
\hat{\mu} = \frac{\beta \Gamma\left( \frac{d}{2}\right) \sin{\left(\frac{\pi d}{\beta} \right)}}{2 \delta \pi^{1+\frac{d}{2}} \langle k_{in} \rangle}.
\end{equation}
This model can be immediately identified as the directed variant of the $\mathbb{S}^d$ model, first introduced in~\cite{allard2024geometric}. It represents a directed extension of the $\mathbb{S}^d$ model originally proposed in~\cite{serrano2008similarity}, along with its equivalent formulation in the hyperbolic plane, known as the $\mathbb{H}^2$ model~\cite{krioukov2010hyperbolic}. Notably, numerous analytical results have been derived for the $\mathbb{S}^1/\mathbb{H}^2$ model, including studies on degree distribution~\cite{serrano2008similarity,krioukov2010hyperbolic,gugelmann2012random}, clustering~\cite{krioukov2010hyperbolic,gugelmann2012random,candellero2016clustering,Fountoulakis2021}, graph diameter~\cite{abdullah2017typical,friedrich2018diameter,muller2019diameter}, percolation~\cite{serrano2011percolation,fountoulakis2018law}, self-similarity~\cite{serrano2008similarity}, and spectral properties~\cite{kiwi2018spectral}. Moreover, this model has been extended to incorporate growing networks~\cite{papadopoulos2012popularity}, weighted networks~\cite{allard2017geometric}, multilayer networks~\cite{kleineberg2016hidden,Kleineberg2017}, and networks with community structure~\cite{zuev2015emergence,garcia-perez:2018aa,muscoloni2018nonuniform}. It also serves as the foundation for defining a renormalization group for complex networks~\cite{garcia-perez2018multiscale,Zheng:2021aa}.

Unlike the DCM, geometry implies that the connection probability is size-independent. In turn, this implies that the reciprocity and clustering are finite, as shown in~\cite{allard2024geometric}. Interestingly, this model undergoes a topological phase transition at the critical inverse temperature $\beta_c = d$~\cite{vanderkolk2022anomalous}. For $\beta > \beta_c$, clustering is finite in the thermodynamic limit, whereas it vanishes below this value. This phase transition is of topological nature and involves the reorganization of cycles in the network, transitioning from being short-range in the clustered phase to long-range in the unclustered one. This transition is accompanied by an anomalous behavior of the entropy per link. From Eq.~\eqref{eq:entropy}, we can compute the entropy as
\begin{equation}
\frac{S}{N\langle k_{in} \rangle} = \frac{2\beta}{d} \left( 1 - \frac{\pi d}{\beta} \cot{\frac{\pi d}{\beta}} \right).
\end{equation}
Unlike standard continuous phase transitions, the entropy per link diverges at the critical temperature from below as
\begin{equation}
\frac{S}{N\langle k_{in} \rangle} \sim \frac{1}{\beta - d},
\end{equation}
whereas it diverges logarithmically at higher temperatures. The origin of this anomalous behavior is due to the fact that the number of available microstates per link at low temperatures is finite, primarily connecting pairs of nodes at bounded distances. However, once the temperature surpasses the critical temperature, the number of available microstates becomes of the order of the number of nodes, as links can now connect pairs of nodes that are arbitrarily far apart.

\subsubsection{Interacting Directed $\mathbb{S}^d$ Model (I-DSM)}

When reciprocal links interact in the directed $\mathbb{S}^d$ model, the strategy applied for the I-DCM, based on using Eq.~\eqref{eq:pijvspij} to relate the connection probabilities in the interacting and non-interacting formulations, cannot be used because $p_{ij}^{ni}$ is independent of the system size and does not approach zero in the thermodynamic limit. The probability of a directed link in I-DSM must be found by substituting Eq.~\eqref{eq:energySd} into Eq.~\eqref{eq:pijcorr}, with $\Delta \varepsilon_{ij} \ne 0$, and imposing sparsity, which leads to new definitions of the chemical potential $\mu$ and the relation between the expected in- and out-degrees of a given node and its in- and out-energies $\varepsilon_{in}$ and $\varepsilon_{out}$. In particular, the connection probability can be written as
\begin{equation}
    p_{ij} = \frac{\chi_{ji}^\beta + e^{-\beta \Delta \varepsilon_{ij}}}{\chi_{ij}^\beta + \chi_{ji}^\beta + \chi_{ij}^\beta \chi_{ji}^\beta + e^{-\beta \Delta \varepsilon_{ij}}},
\end{equation}
where
\begin{equation}
    \chi_{ij} = x_{ij} e^{\varepsilon_{out,i} + \varepsilon_{in,j} - \mu}.
\end{equation}

\begin{figure}[t]
\centerline{\includegraphics[width=0.8\columnwidth]{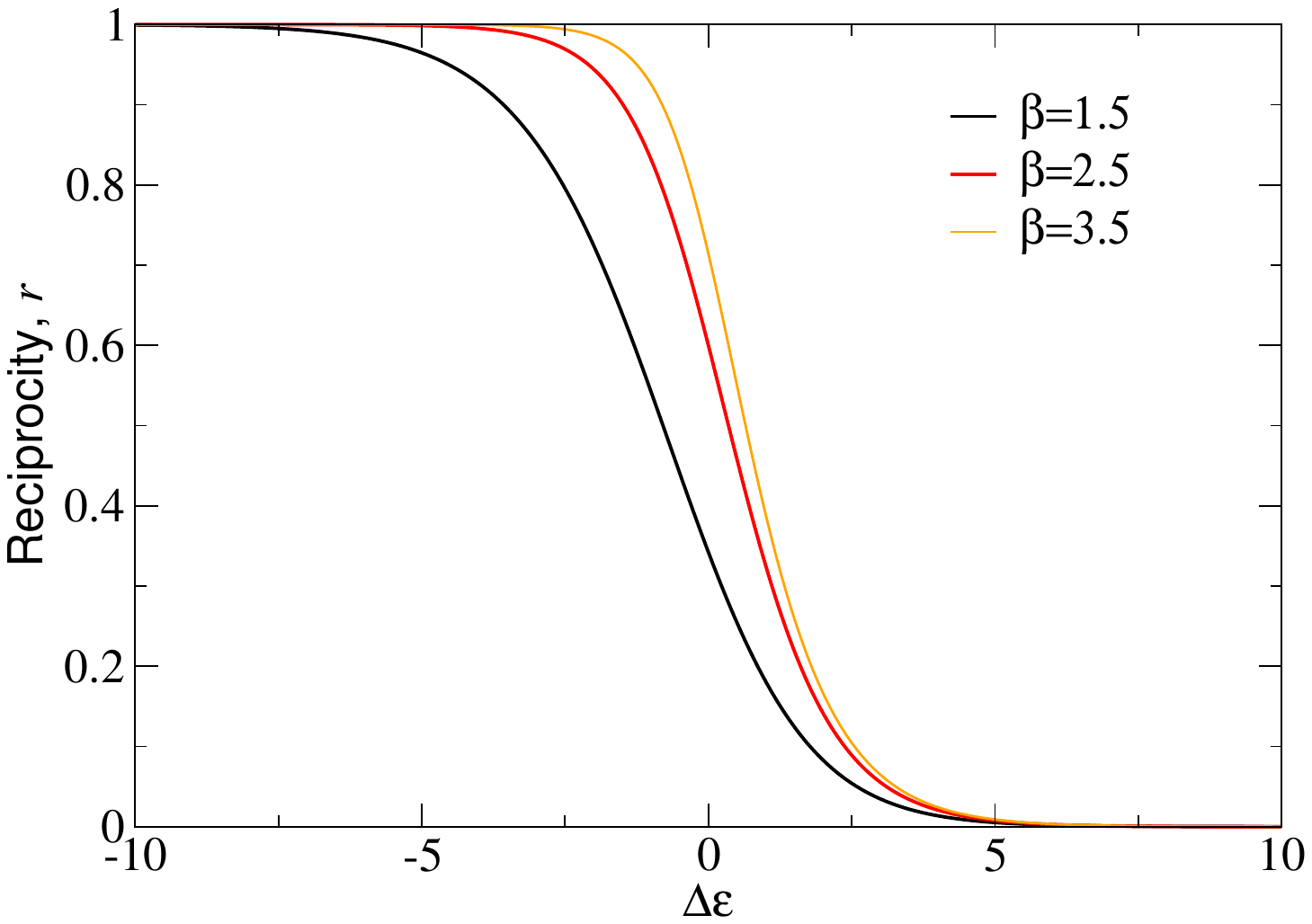}}
\caption{Reciprocity of the interacting directed $\mathbb{S}^d$ model for fully correlated in- and out-energies, as a function of $\Delta \varepsilon$. Different curves correspond to different temperatures $\beta^{-1}$.}
\label{fig:2}
\end{figure}

Using this expression, the average out-degree of a node with in- and out-energies $\varepsilon_{in,i}$ and $\varepsilon_{out,i}$ can be written as
\begin{widetext}
\begin{equation}
    \langle k_{out}(\varepsilon_{in,i}, \varepsilon_{out,i}) \rangle = \delta V_{d-1} e^{d \mu} e^{-d \varepsilon_{out,i}} \int \int e^{-d \varepsilon_{in,j}} \rho(\varepsilon_{in,j}, \varepsilon_{out,j}) d\varepsilon_{in,j} d\varepsilon_{out,j} \int_0^\infty \frac{t^{d-1}(q_{ij}t^\beta + e^{-\beta \Delta \varepsilon_{ij}})}{t^\beta + q_{ij}(1 + t^\beta)t^\beta + e^{-\beta \Delta \varepsilon_{ij}}} dt,
\label{eq:koutfull}
\end{equation}
where $q_{ij} \equiv e^{\varepsilon_{out,j} - \varepsilon_{out,i} + \varepsilon_{in,i} - \varepsilon_{in,j}}$. By integrating Eq.~\eqref{eq:koutfull} over the energies $\varepsilon_{in,i}$ and $\varepsilon_{out,i}$ and equating it to $\langle k_{in} \rangle$, we can obtain the value of the chemical potential $\mu$ from
\begin{equation}
    e^{d \mu} = \frac{\langle k_{in} \rangle}{\delta V_{d-1} \langle e^{-d(\varepsilon_{out,i} + \varepsilon_{in,j})} \int_0^\infty \frac{t^{d-1}(q_{ij}t^\beta + e^{-\beta \Delta \varepsilon_{ij}})}{t^\beta + q_{ij}(1 + t^\beta)t^\beta + e^{-\beta \Delta \varepsilon_{ij}}} dt \rangle},
\end{equation}
\end{widetext}
where the average in the denominator is taken over the random variables $\varepsilon_{in,i}, \varepsilon_{in,j}, \varepsilon_{out,i}, \varepsilon_{out,j}$, and $\Delta \varepsilon_{ij}$. Using a similar approach, the reciprocity becomes
\begin{equation}
    r = \frac{\langle e^{-d(\varepsilon_{out,i} + \varepsilon_{in,j})} \int_0^\infty \frac{t^{d-1}e^{-\beta \Delta \varepsilon_{ij}}}{t^\beta + q_{ij}(1 + t^\beta)t^\beta + e^{-\beta \Delta \varepsilon_{ij}}} dt \rangle}{\langle e^{-d(\varepsilon_{out,i} + \varepsilon_{in,j})} \int_0^\infty \frac{t^{d-1}(q_{ij}t^\beta + e^{-\beta \Delta \varepsilon_{ij}})}{t^\beta + q_{ij}(1 + t^\beta)t^\beta + e^{-\beta \Delta \varepsilon_{ij}}} dt \rangle}.
\end{equation}

Equation~\eqref{eq:koutfull} implies that the average in- or out-degree of a given node depends on both $\varepsilon_{in}$ and $\varepsilon_{out}$, not only on one of them, as is the case for non-interacting fermions. This indicates that computing the degree distributions requires explicitly solving Eq.~\eqref{eq:koutfull}. However, in the particular case of fully correlated $\varepsilon_{in}$ and $\varepsilon_{out}$ and $\Delta \varepsilon_{ij} = \Delta \varepsilon$, the term $q_{ij} = 1$, and the average in- or out-degree becomes a function of $\varepsilon_{in}$ or $\varepsilon_{out}$ separately. Thus, as in the case of non-interacting fermions, we can write $\kappa_{out} \equiv e^{-d\varepsilon_{out}}$ and $\kappa_{in} \equiv e^{-d\varepsilon_{in}}$, with
\begin{equation}
    \mu = -\frac{1}{d} \ln\left(\delta V_{d-1} \tilde{I}(\beta, d, \Delta \varepsilon) \langle k_{in} \rangle \right),
\end{equation}
where
\begin{equation}
    \tilde{I}(\beta, d, \Delta \varepsilon) = \int_0^\infty \frac{t^{d-1}(t^\beta + e^{-\beta \Delta \varepsilon})}{2t^\beta + t^{2\beta} + e^{-\beta \Delta \varepsilon}} dt,
\end{equation}
and the reciprocity becomes
\begin{equation}
    r = \frac{\int_0^\infty \frac{t^{d-1}e^{-\beta \Delta \varepsilon}}{2t^\beta + t^{2\beta} + e^{-\beta \Delta \varepsilon}} dt}{\tilde{I}(\beta, d, \Delta \varepsilon)}.
\end{equation}

Figure~\ref{fig:2} shows the results of the reciprocity in this case as a function of $\Delta \varepsilon$ for different values of $\beta$. Reciprocity converges to 1 in the limit $\Delta \varepsilon \to -\infty$ and approaches zero in the limit $\Delta \varepsilon \to \infty$, as expected. Furthermore, it increases as the temperature rises. Note that the convergence to 1 for very low temperatures and/or highly negative $\Delta \varepsilon$ is only possible in the fully correlated case. In all other cases, the maximum possible value of reciprocity is always less than one.

\section{Conclusions}

The statistical mechanics framework for directed networks introduced in this work treats links as fermionic particles subject to constraints and interactions. This formalism allowed us to describe directed networks within a principled approach that incorporates reciprocity and other structural properties, addressing the limitations of existing models. By leveraging concepts from quantum statistics, our methodology redefines network modeling, shifting the focus from node-centric descriptions to link interactions. Formulating directed networks within a grand canonical ensemble, we demonstrated how the chemical potential and key network features, such as the degree distribution and reciprocity, naturally emerge from the underlying statistical framework. 

The versatility and analytical power of our formalism were illustrated through applications to specific cases, including the Directed Configuration Model and the Directed $ \mathbb{S}^d $ Model. Key results highlighted the influence of interactions on reciprocity and clustering. In the non-interacting formulations, reciprocity vanishes in the thermodynamic limit, whereas in the interacting models, the framework supports a tunable reciprocity that remains finite under specific conditions. The inclusion of a geometric component in the $ \mathbb{S}^d $ model further showcased how spatial constraints shape the emergent properties of the network. This framework bridges theoretical advances with empirical applicability, providing a robust toolset for analyzing real-world directed networks. Additionally, it paves the way for exploring dynamical processes on directed topologies and designing models that better reflect the intricate balance of directed interactions. Future work could extend these principles to multilayer, temporal, or weighted networks, offering a deeper understanding of complex systems.

\section{Acknowledgments}

We acknowledge support from: Grant TED2021-129791B-I00 funded by MCIN/AEI/10.13039/501100011033 and by  ``European Union NextGenerationEU/PRTR''; Grant PID2022-137505NB-C22 funded by MCIN/AEI/10.13039/501100011033 and by ``ERDF A way of making Europe''; Generalitat de Catalunya grant number 2021SGR00856. M. B. acknowledges the ICREA Academia award, funded by the Generalitat de Catalunya.


%

%

\end{document}